\documentclass{aa}
\begin{document}

   \thesaurus{06     
              (13.09.4;  
               13.09.1;  
               11.19.2;  
               11.09.4;  
               09.13.2)} 

   \title{Gas cooling within the diffuse ISM of late--type galaxies
\thanks{Based on observations with the Infrared Space
Observatory (ISO),
an ESA project with instruments funded by ESA member states
(especially the PI countries: France, Germany, the Netherlands
and the United Kingdom) and with the participation of ISAS and
NASA.}}


   \author{D. Pierini
          \inst{1,2}
          \and
          J. Lequeux\inst{3}
          \and
          A. Boselli\inst{4}
          \and
          K.J. Leech\inst{5}
          \and
          H.J. V\"olk\inst{2}
          }

   \offprints{D. Pierini}

   \institute{Ritter Astrophysical Research Center, University of Toledo,
              2801 West Bancroft, Toledo, OH 43606\\
              email: pierini@ancona.astro.utoledo.edu\\
         \and Max-Planck-Institut f\"ur Kernphysik,
              Saupfercheckweg 1, D-69117 Heidelberg\\
              email: Heinrich.Voelk@mpi-hd.mpg.de\\
         \and DEMIRM, Observatoire de Paris,
              61 Av. de l'Observatoire, F-75014 Paris\\
              email: James.Lequeux@obspm.fr\\
         \and Laboratoire d'Astronomie Spatiale,
              BP8 Traverse du Syphon, F-13376 Marseille\\
              email: Alessandro.Boselli@astrsp-mrs.fr\\
         \and ISO Data Centre, Astrophysics Division, ESA Space Science Dept.,
              P.O. Box 50727, S-28080 Madrid\\
              email: kleech@iso.vilspa.esa.es
             }

   \date{Received ..., 2000; accepted ..., ...}

   \maketitle

   \begin{abstract}

We combine observations of spiral galaxies in the [CII]
line at 158 $\mu$m, made with the Long Wavelength Spectrometer aboard ISO,
with previous data from the Kuiper Airborne Observatory to study the origin
of this line, which is the main coolant of the interstellar medium
at relatively low temperatures.
We also use HI and CO(1-0) observations of these galaxies
and estimate the respective line fluxes in the same beam
as the [CII] observations.
We confirm the existence of a linear relation between
the [CII] line intensity and the CO(1-0) line intensity,
that we extend to intrinsically fainter galaxies.
The dispersion around this relation is significant and due to variations
in the far-UV flux, thus in the star formation rate.
We find that for the least active galaxies of our sample, in terms
of star formation, the rate of [CII] line emission
per interstellar hydrogen atom is similar to that in the Solar neighbourhood.
For those galaxies, most of the [CII] line emission comes
probably from the diffuse cold atomic medium.
In more active galaxies, considered globally, the average [CII]
line emission is dominated by dense photodissociation regions
and to some extent by the warm ionized diffuse medium.
This is true in the central region of many spiral galaxies, and probably
even in the interarm regions of the most actively star--forming ones.

      \keywords{Infrared: ISM: lines and bands --
                Infrared: galaxies --
                Galaxies: spiral --
                Galaxies: ISM --
                ISM: molecules
               }

   \end{abstract}

\section{Introduction}

The [CII]($\rm ^2 P_{3/2}~-~^2 P_{1/2}$)
($\lambda = 157.7409~\mu$m) line of singly ionized carbon
is the dominant cooling line in the diffuse Interstellar Medium (ISM)
at gas temperatures $\rm T \le 8 \times 10^3~K$ (Dalgarno \& McCray 1972;
Tielens \& Hollenbach 1985; Wolfire et al. 1995).
Excitation of the $\rm ^2 P_{3/2}$ level of $\rm C^{+}$ is due
to inelastic collisions either with neutral hydrogen atoms and molecules
or with electrons (Dalgarno \& McCray 1972; Stacey 1985;
Kulkarni \& Heiles 1987).
Neutrals dominate over electrons if the fractional ionization
$X_{\rm e} = n_{\rm e} / (n_{\rm H} + 2 n_{\rm H_2})$ is smaller
than $\rm 10^{-3}$ (Dalgarno \& McCray 1972).

Heating of the interstellar gas is mainly due to photo--electrons emitted
by 3-D dust grains and Polycyclic Aromatic Hydrocarbons (PAHs) submitted
to ultraviolet radiation from stars (Bakes \& Tielens 1994), both in
the diffuse ISM (Wolfire et al. 1995) and in Photodissociation Regions (PDRs),
at the interfaces between molecular clouds and HII regions
(Tielens \& Hollenbach 1985; Bakes \& Tielens 1998).
The photoelectric effect is due essentially to photons
with $\rm 6 \le h \nu < 13.6~eV$.
In the field of a galaxy, this radiation is dominated by B3 to B0 stars
with $\rm 5 \leq M \leq 20~M_{\sun}$ (e.g. Xu et al. 1994) but of course
hotter, more massive stars can also contribute locally.

Earlier studies of the inner regions of gas--rich spirals
and starburst galaxies (Crawford et al. 1985; Stacey et al. 1991;
Carral et al. 1994) have found that the [CII] line intensity,
$I_{\rm C II}$, is typically a few $\rm \times~10^{-3}$ of the total far-IR
continuum dust emission, $I_{\rm FIR}$, within the same region.
For these galaxies, the previous authors have also found a linear relation
between $I_{\rm C II}$ and the $\rm ^{12}CO(J:1 \rightarrow 0)$
($\lambda = 2.6$ mm) line intensity, $I_{\rm CO}$, with a scatter
slightly larger than one order of magnitude.
Conversely, the correlation with atomic hydrogen is poor.
This led them to the conclusion that PDRs are the main sources
of the [CII] emission on a galaxy scale (see also Wolfire et al. 1989).

However space--resolved [CII] observations of the large spiral
galaxy NGC\,6946 by Madden et al. (1993) showed that this emission originates
also in the diffuse ISM (mostly associated to the interarm regions).
More precisely, 5\% of the total line luminosity in this galaxy comes from
its nucleus, 20\% from the spiral arms with their HII regions
and PDRs, and 75\% from an extended component, the fractional contribution
of which increases with radius.
This result is reproduced by the model calculations of Sauty et al. (1998).

Unfortunately resolved galaxies on which such a detailed study can be
conducted are not numerous.
We will thus present a discussion of the relative contribution of the diffuse
medium and of dense PDRs in spiral galaxies based on a relatively large sample
of galaxies observed in the [CII] line at their central position,
which add to the previous galaxies observed with the Kuiper Airborne
Observatory (KAO).

A substantial number of galaxies have been observed in the [CII]
line with the Long Wavelength Spectrometer (LWS) (Clegg et al. 1996)
aboard ISO (Kessler et al. 1996).
Malhotra et al. (1997) observed 30 normal star--forming galaxies,
but unfortunately do not give data for individual galaxies.
Smith \& Madden (1997) report results on 5 low blue luminosity spirals
in the Virgo cluster, Braine \& Hughes (1999) measurements of the normal
spiral NGC\,4414 and Lord et al. (1996) observations of the peculiar galaxy
NGC\,5713.
Leech et al. (1999) detected 14 out of 19 Virgo cluster spiral galaxies
with relatively low massive star--formation (SF) activity.
All the latter galaxies are in general less active in terms of
star--formation than the galaxies observed with the KAO.

Using the sample of Leech et al. (1999) together with the KAO sample,
Pierini et al. (1999 - hereafter referred to as P99) have shown that
the [CII]-to-CO line intensity ratio, $I_{\rm C II}/I_{\rm CO}$,
is a reliable tracer of the global mass--normalized massive star--formation
rate (SFR) of normal spiral galaxies, as measured by the equivalent width
of the H$\alpha$ line (H$\alpha$ EW), which is proportional to the ratio
of the stellar Lyman continuum to the stellar red continuum.

They also observed a non--linear dependence of $I_{\rm C II}/I_{\rm FIR}$
on the H$\alpha$ EW, where $I_{\rm FIR}$ is determined as in Helou et al.
(1985).
The so-called ``normal star--forming'' galaxies with H$\alpha$ EW $\ge$
10 \AA~have values of $I_{\rm C II}/I_{\rm FIR} \sim 4 \times 10^{-3}$,
while the ``quiescent'' galaxies with H$\alpha$ EW $<$ 10 \AA, mainly
identified with early--type spirals, show a continuous decrease
in $I_{\rm C II}/I_{\rm FIR}$ with decreasing H$\alpha$ EW.
P99 interpreted the latter behaviour in terms of a dominant
``[CII]--quiet'' component of the far-IR emission,
due to dust heating by relatively low mass stars, which do not emit
enough non--ionizing far-UV radiation to produce a substantial photoelectric
effect (see e.g. Xu et al. 1994).
This scenario has been recently invoked to interpret the [CII]
emission of early--type galaxies (Malhotra et al. 2000).

Assuming that the [CII] emission characteristics of diffuse
and compact Galactic sources, empirically established by Nakagawa et al.
(1998), are the same as those of the normal galaxies in their sample, P99
finally showed that the diffuse components of the ISM are at least
as important sources of gas cooling in these galaxies as are compact regions,
in agreement with recent arguments in the literature (Madden et al. 1993;
Heiles 1994; Bennett et al. 1994; Dwek et al. 1997; Smith \& Madden 1997;
Sauty et al. 1998).

Here we want to discuss the latter conclusion in more detail,
taking into account the gas content in its molecular and atomic phases
at different galactocentric radii.
We will use the total sample of 42 late--type galaxies observed by Stacey
et al. (1991) (the KAO sample), Smith \& Madden (1997) and Leech et al. (1999).
The sample and the related data--base are discussed in Sect. 2.
Our observational results are shown in Sect. 3.
The discussion and the summary of our conclusions are contained in Sect. 4
and 5, respectively.

\section{The galaxy sample and complementary data}

\subsection{The Sample}

The ISO Guaranteed Time program ``VIRGO'' combines studies of a deep
optically complete, volume--limited sample of spiral, irregular
and blue compact dwarf galaxies, selected from the Virgo Cluster Catalogue
(VCC) of Binggeli et al. (1985).
117 member galaxies later than S0 and with $\rm B_T \le 18~mag$ were selected
in the core and at the periphery of the cluster (Boselli et al. 1997).
62 targets were observed with ISOPHOT at 60, 100 and 160 $\mu$m
and a complete sample of 95 targets was observed with ISOCAM through
the LW2 and LW3 filters centered respectively at 6.75 and 15 $\mu$m.
Due to time constraints, only 19 normal spiral galaxies
(18 with $\rm B_T \le 12.3~mag~+$ NGC\,4491, with $\rm B_T = 13.4~mag$)
could be observed with the LWS in the [CII] line
by Leech et al. (1999).
Fourteen galaxies were detected in the [CII] line
with a signal-to-noise ratio above $\rm 3 \sigma$; upper limits were set
to the other 5.
Assuming a distance of 21~Mpc from Virgo, the 70\arcsec~half power
beam--width (HPBW) of the LWS corresponds to $\sim$ 7~kpc and encompasses
at least one exponential disk scale--length, with the exception of
the galaxies with optical major axis larger than 5 arcmin.
5 other VCC spiral galaxies with $\rm 12.5 \le B_T \le 13.9~mag$,
observed with the LWS in the [CII] line by Smith \& Madden (1997)
are added to this sample.
We call the total sample the ISO sample.
This sample contains spirals with a large dynamic range in mass--normalized
massive star--formation rate (SFR) ($\rm 0 \le H \alpha~EW \le 71~\AA$)
and in morphological type (from S0/a to Sd).

H$\alpha$+[NII] equivalent widths are available for 20
out of these 24 galaxies either from CCD imaging (Boselli et al.
in preparation; Gavazzi et al. in preparation) or, in a few cases,
from long--slit spectroscopy of their central regions (Kennicutt \& Kent 1983).
The latter equivalent widths represent fairly well the whole mass--normalized
massive star--formation rate, since the 3\arcmin--7\arcmin~slits used
by Kennicutt \& Kent (1983) cover almost the entire galaxy.
19/24 galaxies have detections of the CO(1-0) line at their central position
or upper limits.
13/24 galaxies have HI ($\lambda =$ 21 cm) surface density
profiles deprojected onto the galaxy plane from Warmels (1988), while 17/24
have observed HI fluxes listed in de Vaucouleurs et al.
(1991 - RC3).

The parameters of the ISO sample are given in Table 1 as follows: \newline
Column 1: NGC and VCC numbers; \newline
Column 2: morphological type (RC3 number); \newline
Column 3: logarithm of the length of major axis from the RC3; \newline
Column 4: logarithm of the axial ratio from the RC3; \newline
Column 5: H$\alpha$+[NII] equivalent width
(H$\alpha$ EW for short); \newline
Column 6: CO(1-0) line intensity (in the main beam temperature scale)
at the central position of the galaxy, interpolated/extrapolated at/to
the 70\arcsec~LWS resolution as described in Sect. 2.2; \newline
Column 7: average neutral hydrogen column density in the 70\arcsec~LWS beam,
$N$(HI), derived as shown in Sect. 2.2; \newline
Column 8: observed [CII] line flux; \newline
Column 9: references for the CO(1-0) and HI data. \newline
In the footnotes, we give the FWHM of the CO distribution
of individual galaxies, as determined in Sect. 2.2.

Data for the KAO sample from Stacey et al. (1991) (hereafter referred to
as S91) are given in Table 2.
These authors give average [CII] line surface brightnesses
for the central regions of 18 galaxies of their sample, observed with a beam
of 55\arcsec~FWHM, from which we derive line fluxes comparable to those
of Table 1.

S91 also report central measurements of the CO(1-0) line emission
(roughly at the same angular resolution as their KAO measurements)
for the whole sample and measurements in the HI line emission
(at different resolutions) for 16/18 galaxies.
HI flux measurements are listed in the RC3 for 17/18 objects.
Unfortunately, H$\alpha$ EWs are available only for a third of the KAO sample,
from the 3\arcmin~slit spectroscopy of Kennicutt \& Kent (1983) and, therefore,
they are poor approximation to the integrated value of the mass--normalized
massive star--formation rate.

In Table 2 we give the parameters of the KAO sample galaxies as follows:
\newline
Column 1: galaxy name; \newline
Column 2: morphological type from the RC3; \newline
Column 3: logarithm of the length of major axis from the RC3; \newline
Column 4: logarithm of the axial ratio from the RC3; \newline
Column 5: H$\alpha$+[NII] equivalent width (H$\alpha$ EW for short); \newline
Column 6: CO(1-0) line intensity (in the main beam temperature scale)
at the central position of the galaxy, approximately at the 55\arcsec~KAO
resolution; \newline
Column 7: average neutral hydrogen column density in the 55\arcsec~KAO beam,
$N$(HI), derived as shown in Sect. 2.2; \newline
Column 8: observed [CII] line flux; \newline
Column 9: ``sb'' and ``gr'' refer to starburst and gas--rich galaxies,
respectively, i.e. to galaxies with dust temperatures in excess of 40 K (sb)
and less than 40 K (gr), according to S91.

We divide the total ISO sample into quiescent and normal star--forming
galaxies according to the values of H$\alpha$ EW $<$ 10 \AA~and $\ge$ 10 \AA,
respectively, when H$\alpha$ measurements are available.

\subsection{Data analysis}

A major difficulty in our analysis is the lack of homogeneous
[CII], CO and HI data obtained
at the same angular resolution.
We will try to remedy this situation by interpolating or extrapolating
the CO and HI measurements to the resolution
of the [CII] observations, and will consider mainly
surface brightnesses in the KAO or LWS beam.

While S91 give CO line fluxes obtained at angular resolutions rather similar
to their beam, the situation is more heterogenous for the ISO sample.
CO line fluxes of Virgo galaxies have been measured at different
angular resolutions by Young et al. (1985), Stark et al. (1986),
Kenney \& Young (1988), Boselli et al. (1995) and Smith \& Madden (1997).
Fortunately, the surveys of Stark et al. (1986) (100\arcsec~FWHM)
and Kenney \& Young (1988) (45\arcsec~FWHM) have 28 objects in common,
11 of which belong to our ISO sample.

In Fig. 1 we reproduce the plot of the central CO line intensity
(in $\rm K~km~s^{-1}$ in the main beam temperature scale)
for these 28 objects as observed by these two radiotelescopes,
from Kenney \& Young (1988).
Detections and upper limits are considered.
The filled circles represent the 11/28 galaxies in common with
the present study.
We see that the representative points scatter nicely between a line
of slope 1 expected for galaxies with a CO distribution much more extended
than both beams, and a line of slope (45/100)$^2$ expected for distributions
much smaller than both beams.
In particular, it is inferred that the CO distribution of NGC\,4654
is extended, while the ones of NGC\,4178, NGC\,4394 and NGC\,4450
are point--like.
Therefore, for NGC\,4654, the average CO brightness in the 70\arcsec~ISO-LWS
beam is derived from the average of the measurements at 45\arcsec~and
100\arcsec~resolution.
Instead, for NGC\,4178, NGC\,4394 and NGC\,4450, the average CO brightness
in the ISO-LWS beam is derived by dividing the CO line flux by the ISO-LWS
beam area.

For the remaining 6 ISO galaxies observed and detected in CO
at these two different angular resolutions, we apply a different method.
Assuming that both antenna beams are circular and gaussian and that
the CO brightness is gaussian with the same shape as the optical galaxy
(given by the RC3), we can derive first the FWHM of the CO distribution
along the major axis from the observed flux ratio, then the average
CO brightness in the ISO-LWS beam (also assumed to be circular gaussian).
The values of the FWHM of the CO distribution along the major axis
of these 6 galaxies are given in the footnotes of Tab. 1.

For the galaxies NGC\,4178, NGC\,4450 and NGC\,4569 the data
of Boselli et al. (1995) (34 and 43\arcsec~FWHM) are preferred to the ones
of Kenney \& Young (1988) since their accuracy is much better.

For the galaxy NGC\,4698 only upper limits are available, so that we estimate
a mean upper limit to its average CO brightness in the LWS beam.

For the remaining 7 ISO galaxies with available CO measurements, but taken
only at one radiotelescope, we assume that the ratio of the CO
to the optical size is the same as the average one determined from
the 27/28 galaxies observed by both Stark et al. (1986) and Kenney \& Young
(1988) with available measurements/upper limits, and calculate
the average CO brightness in the LWS beam.
This method introduces some uncertainty since the average ratio
of the CO to the optical size (0.32) has a rms dispersion of 0.30
for the 27 galaxies.
We find that this ratio does not depend on the morphological type
of the galaxy.
The previous 7 ISO galaxies represent 40\% of the ISO galaxies
with estimated average CO brightness in the LWS beam
and 20\% of the total sample galaxies in Fig. 4.

The HI data--set is more heterogeneous.
S91 report average HI column densities from maps obtained
at very different angular resolutions with respect to the KAO beam.
However, 17/18 galaxies of the latter sample have observed
HI fluxes listed in the RC3.
Although the spatial information is not available, the latter measurements
constitute at least a homogeneous data--set and are preferred to those ones
adopted by S91.

17/24 galaxies of the ISO sample have observed HI fluxes
listed in the RC3 as well.
In addition, 14 galaxies of the ISO sample have HI surface
density profiles determined by Warmels (1988) after correction
for the inclination of the galaxy.
For these 14 ISO galaxies, we use these data to calculate the average
HI column density $N$(HI) in the 70\arcsec~LWS beam by correcting backwards
the Warmel's data into apparent surface densities, not deprojected.

For the 2 ISO galaxies (NGC\,4293 and NGC\,4522) and the 17 KAO galaxies
with only a HI flux, we derive a ``hybrid'' average surface brighness
in the LWS or KAO beam by assuming that the distribution of HI is gaussian
with a FWHM equal to the blue light dimension as given in the RC3.
Again, our assumption is reasonable but introduces some uncertainty.
Such ``hybrid'' surface brightnesses are admittedly crude
but there is no alternative.

For the 14 ISO galaxies with data from Warmels, the average ratio
between the average column densities in the ISO-LWS beam derived from
the HI surface brightness profiles and those estimated through
the ``hybrid'' surface brightness method is 1.5, with a rms dispersion of 0.9.
The difference in these average surface brightnesses is not large enough
to justify a further correction.
Although the dispersion is large, we are confident that the conclusions
derived in the following sections are not biased by the use of these crude
HI surface brightnesses.

Note that in all the previous discussion we have eliminated the interacting
galaxy NGC\,4438, which is a special case.

\section{Results}

Due to our observational limitations in terms of angular resolution,
we find useful to quantify the fraction of a galaxy observed either
with the KAO or with the ISO-LWS in terms of a coverage factor CF.
This coverage factor is the ratio between the area of the observing beam
and the projected optical galaxy area as defined in the RC3.

Fig. 2 shows a plot of the observed central [CII] line intensity
vs. the CF of the 42 galaxies of our sample.
It confirms that the KAO probed mainly the innermost regions
of both gas--rich and starburst galaxies.
This is in part because the galaxies of this sample are relatively nearby.
The ISO galaxies are further away and as a consequence the LWS sampled
larger areas of the galaxies.
We note that the 5 ISO galaxies with [CII] line upper limits
fall well within the range of values of CF of the galaxies detected
with the LWS, so that we are led to the conclusion that their faintness
is not due to a particularly reduced spatial sampling but 
to a genuine deficiency of [CII] line emission.
The Hubble type of these 5 undetected galaxies is either S0/a or Sa;
galaxies of these types have generally a lower SFR than galaxies
of later type, so the faintness of their emission is not unexpected.

In Fig. 3, we plot as a function of the CF the average column densities
of molecular hydrogen in the beam, $2 N({\rm H_2})$, (in terms of
HI nuclei) and those of the neutral atomic hydrogen,
also in the beam, $N$(HI), both not deprojected
from the inclination of the galaxy (panels a and b, respectively).
$2 N({\rm H_2})$ is derived from $I_{\rm CO}$ as follows (Digel et al. 1995):
\newline
$2 N({\rm H_2})~=~2.12~(\pm 0.28) \times 10^{20}$ $I_{\rm CO}$, \newline
where $2 N({\rm H_2})$ is in units of $\rm cm^{-2}$
and $I_{\rm CO}$ is in units of $\rm K~km~s^{-1}$.

This value of the $\rm H_2$ column density-to-CO line intensity ratio
has been derived in Orion and we use it for all our galaxies,
although it may depend on metallicity and on the ISRF intensity
(cf. Israel 1997 and references therein).
Its uncertainty is equivalent to 13\% of $2 N({\rm H_2})$.

We arbitrarily assume that both $N$(HI) and $2 N({\rm H_2})$ are affected by
a $1 \sigma$ uncertainty of 20\%.

For both starburst and non--starburst galaxies, the inner regions are richer
in molecular hydrogen than the outer regions (cf. Young \& Knezek 1989).
This is not the case for atomic hydrogen.
We note that the ISO early--type spiral galaxies have the lowest values
of $N$(HI) (cf. Tab. 1 and 2), in agreement with Solanes et al. (2000).

The innermost regions of the KAO galaxies are rich in molecular gas,
as witnessed by their high molecular-to-atomic hydrogen
column density ratio ($2 N({\rm H_2})$/$N$(HI) $\ge$ 4),
or more exactly by their high CO/HI line intensity ratio.
Therefore, we might reasonably expect that, in these regions, a relevant part
of the gas cooling via [CII] line emission comes from PDRs associated
to molecular clouds where star--formation activity may be more or less active.
In particular, this may account partly for the linear relationship between
$I_{\rm C II}$ and $I_{\rm CO}$ found by S91 for their starburst galaxies.

In Fig. 4 we show this relationship for the 36 galaxies of our total sample
that are detected at the $\rm 3 \sigma$ level in at least one
of the two lines, except for NGC\,4596.
The long-- and short--dashed lines represent the two characteristic
values of $I_{\rm C II}$/$I_{\rm CO}$ given by S91 for their ``warm dust
galaxies'' (i.e. starburst galaxies) and ``cold dust galaxies''
(i.e. normally active gas--rich galaxies), respectively.
The ISO observations extend this relation to line fluxes fainter
by 1.5 orders of magnitude.
The overall linear relation remains, although with a probably more
significant rather large scatter, even though part of it is due
to the uncertainties of our data analysis (cf. Sect. 2.2).
The interpretation given by P99 for this scatter is that
$I_{\rm C II}$/$I_{\rm CO}$ is proportional to the strength
of the far-UV radiation field, as represented by the $\rm H \alpha$
equivalent width.
This is confirmed by the model calculations of PDRs of various densities
and of the diffuse ISM presented by Kaufman et al. (1999), their Fig. 9.

In Fig. 4, three ISO galaxies (NGC\,4178, NGC\,4222 and NGC\,4299)
have exceptionally high values of $I_{\rm C II}$/$I_{\rm CO}$, typical
of starburst galaxies, while their $I_{\rm C II}$ are slightly lower than
for galaxies of the same Hubble type and with similar CFs.
NGC\,4222 has a particularly high inclination but small H$\alpha$ EW
(cf. Tab. 1), so that its ``anomalous'' behaviour in Fig. 4
is probably spurious.
In fact, the latter vanishes when correcting $I_{\rm C II}$ for inclination
(cf. Wolfire et al. 1989), while the ``anomalous'' behaviours
of NGC\,4178 and NGC\,4299 are confirmed.
NGC\,4178 is an HII galaxy, with some giant HII
regions in its inner part (Boselli et al. in preparation), probed by LWS.
Its ``anomalous'' behaviour was not noted by P99 due to the different value
of $I_{\rm CO}$ adopted by these authors but is not due to the assumption
that its CO morphology is point--like (Sect. 2.2).
In fact, $I_{\rm C II}$/$I_{\rm CO}$ decreases by only a factor of -0.18
in logarithm under the assumption that the CO surface brightness distribution of NGC\,4178 is partially resolved at 43\arcsec~resolution.
NGC\,4178 and NGC\,4299 have bluer B-V colour indices (Gavazzi, private
communication) than the other ISO galaxies of the same Hubble type
considered here.
In addition, they have also exceptionally low values
of $2 N({\rm H_2})$/$N$(HI).
These behaviours lead us to the conclusion that NGC\,4178 and NGC\,4299
have intrinsic particularly intense ISRFs (cf. Israel 1997
and references therein) and/or intrinsic low metallicity
(cf. Smith \& Madden 1997).

\section{Discussion: [CII] line emission by the diffuse ISM}

The correlations presented above have not yet shed light about
the [CII] line emission by the diffuse ISM.
Before entering this discussion, it is interesting to summarize
what we know about our own Galaxy.

The diffuse ISM in our own Galaxy is thought to consist
(Kulkarni \& Heiles 1987; Heiles 1988, 1994; Reynolds 1993;
Heiles et al. 1996) of: i) a Cold Neutral Medium (CNM), with a temperature
T of about 80~K, a hydrogen density $n_{\rm H}$ of about $\rm 90~cm^{-3}$
and a fractional ionization $X_{\rm e}$ within a factor of 3
of $\rm \sim 6~\times~10^{-4}$; ii) a Warm Neutral Medium (WNM),
with $\rm T~\sim 8000~K$, $n_{\rm H}~\le \rm 1~cm^{-3}$
and $X_{\rm e}~\sim 3~\times~10^{-2}$; iii) a Warm Ionized Medium
(WIM), with $\rm T~\sim 8000~K$, $n_{\rm H}~\le \rm 1~cm^{-3}$
and $X_{\rm e} \rm ~\ge 0.75$.
The distinction between the WNM and WIM is not completely clear,
and it may be that they are partly one and the same thing.

In our Galaxy, the high latitude [CII] line emission
is associated to the CNM (Bennett et al. 1995) while, in the inner regions
of the disk, the WIM has been proposed as the main source
of the [CII] line emission (Heiles 1994).
At high Galactic latitudes, no prominent OB associations are found,
while the opposite is true for the inner regions (Bronfman et al. 2000),
favoring the WNM and WIM.
For the local Galaxy near the Sun, Bennett et al. (1995) find
a [CII] line emission per interstellar hydrogen nucleus
of $\rm \sim 2.5 \times 10^{-26}~ergs~s^{-1}~H^{-1}$, while Madden et al.
(1993) find a value of $\rm \sim 1.6 \times 10^{-25}~ergs~s^{-1}~H^{-1}$
for the diffuse ISM of NGC\,6946, an active star--forming galaxy.
It is interesting to compare these observations to the results of the model
calculations of Wolfire et al. (1995) which relate to the local Galactic ISM:
they predict a cooling rate per hydrogen nucleus (dominated by
the [CII] line) from 2.6 to
$\rm 6.6 \times 10^{-26}~ergs~s^{-1}~H^{-1}$ in the CNM and from 0.3 to
$\rm 0.8 \times 10^{-26}~ergs~s^{-1}~H^{-1}$ in the WNM,
depending on the exact values of the physical parameters of the medium.
Moreover, they estimate a value of about
$\sim 2.7 \times 10^{-26}~(P/k)_3~\rm ergs~s^{-1}~H^{-1}$ in the WIM,
where $(P/k)_3$ is the WIM pressure in units of $\rm 10^3~K~cm^{-3}$.
Their lower value for the CNM compares favorably with the observation
of Bennett et al. (1995).
An increase of the UV flux would raise the cooling rate from the CNM
appreciably, but this flux should be higher by considerably more
than one order of magnitude to reach the cooling rate of the interarm medium
of NGC\,6946 (see Fig. 3 of Kaufman et al. 1999 and Table 5
of Wolfire et al. 1995).

In Fig. 5, we plot the quantity
$\Lambda~=~\rm 4 \pi$ $I_{\rm C II}$/$N$(HI) for our galaxies
as a function of the coverage factor CF.
If the diffuse ISM was the only source of the [CII] line
emission, this quantity should be in the same range of values than the numbers
given above for the diffuse ISM of our Galaxy and most probably lower
or equal to that for the interarm medium of NGC\,6946.
Any contribution of the dense PDRs would raise $\Lambda$.
Fig. 5 shows that when the beam covers most of the galaxy (CF$\gg$0.1)
$\Lambda$ is indeed in this range, while it increases strongly when only
the central regions are sampled.
This demonstrates the importance of [CII] line emission
by the diffuse ISM in non--starburst galaxies when one considers
the whole galaxy.
Moreover the above discussion shows that the CNM dominates the emission
in normal circumstances.

We note that the normal star--forming galaxy NGC\,4178 has a relatively low
value of $\Lambda$.
This is not unexpected if NGC\,4178 is indeed a low metallicity galaxy
(Wolfire et al. 1995).

We present in Fig. 6 the same plot as in Fig. 5 but considering this time
the line emission per total hydrogen atom whether included or not in molecules
($\Lambda^{\prime}~=~4 \pi$ $I_{\rm C II}$/($N$(HI)+$2 N(\rm H_2)$).
Since the contribution of H$_2$ to the total gas is minor at the scale
of a whole galaxy, both plots are very similar at large values of CF.
This is not the case in the inner regions which correspond to small values
of CF, but the [CII] emission per total H is not very much
higher there than for entire galaxies.
This conclusion is however uncertain because the conversion factor between
the CO(1-0) line intensity and the column density of $\rm H_2$
is very poorly known in the center of galaxies.
As expected, galaxies with more active SF have on average a higher value
of the cooling rate per hydrogen nucleus.
This confirms that the total [CII] line emission
per hydrogen nucleus increases with the strength of the local far-UV field.

From our own data alone it is not directly possible to infer the fraction
of the [CII] line luminosity which comes from the diffuse
ISM and from the PDRs respectively, because these data are spatially averaged
at our limited angular resolution.
However we note that the lowest values of $\Lambda$ on Fig. 5 are similar
to the local value determined by Bennett et al. (1995)
for the solar neighborhood.
Because the [CII] line intensity is relatively insensitive
to the far-UV flux density, this local value should be typical for relatively
quiescent galaxies.
Consequently there is little room left for the contribution of PDRs
to the [CII] line luminosity for the galaxies
with the lowest values of $\Lambda$ in Fig. 5.
On the other hand, the high value quoted for the interarm ISM of NGC\,6946
by Madden et al. (1993) would require a very high far-UV radiation density,
perhaps unrealistic, if it is assumed to be due to a cold neutral medium
alone, (Wolfire et al. 1995), and therefore it is likely that there is already
a contribution of dense PDRs at such high values of $\Lambda$.
The statistical results of P99 are consistent with this estimate.
Since they used different local Galactic data and a different parametrization,
our result is also complementary with P99.

In the {\it inner} regions of the sample galaxies, dense PDRs always seem
to dominate as shown by the high values of $\Lambda$ in Fig. 5, even though
the WIM could give a non-negligible contribution (Heiles 1994).

\section{Conclusions}

We have made or collected observations of the central regions
of a relatively large sample of spiral galaxies
in the [CII]($\rm ^2 P_{3/2}~-~^2 P_{1/2}$)
($\lambda = 158~\mu$m) gas cooling line.
This sample, mainly observed with the Long Wavelength Spectrometer aboard ISO,
enlarges a previous sample observed with the Kuiper Airborne Observatory
and contains on average more quiescent, fainter galaxies.

For the total KAO $+$ ISO-LWS sample of 42 galaxies, we have collected
measurements of the atomic and molecular gas, and estimated
their average column densities in the same beam as the [CII]
line observations.
This beam covers various fractions of the galaxy disks according to
their distance and linear size.

Our results are more detailed than but complementary to those of Pierini
et al. (1999).
We confirm the existence of an overall linear relation between
the [CII] line intensity, $I_{\rm C II}$,
and the CO(1-0) line intensity, $I_{\rm CO}$, over almost 2.5 orders
of magnitude in both quantities.
The significant large scatter around this relation is due to the variation
of the far-UV radiation field and is in agreement with model calculations
by Kaufman et al. (1999).

We find that the [CII] line emissivity per interstellar
atomic hydrogen nucleus, $\Lambda$, ranges between
$\rm \sim 2.6 \times 10^{-26}$
and $\rm \sim 4 \times 10^{-25}~ergs~s^{-1}~H^{-1}$ in galaxies
where the observing beam cover at least 10\% of the optical surface
of the galaxy.
The lowest observed values of $\Lambda$ are similar to the value obtained
by Bennett et al. (1995) for the diffuse ISM near the Sun,
where the atomic cold neutral medium (CNM) dominates the emission.
For the corresponding galaxies, it is likely that most
of the [CII] line emission also comes from the CNM.
The dense Photodissociation Regions plus the Warm Ionized Medium dominate
in more active star--forming galaxies, even including their extended regions,
and they do so in the inner parts of all spiral galaxies.

Finally, we find that the cooling rate per hydrogen nucleus,
when both the molecular and the atomic phases are taken into account,
increases moderately with the star--formation rate, although it is difficult
to give a quantitative result because of the uncertainty in the amount
of $\rm H_2$ in regions of intense star--formation.
\vskip 1.0truecm
\begin{acknowledgements}

   We are grateful for the support of this work to the
   \emph{Deutsche Agentur f\"ur Raumfahrt Angelegenheiten},
   through \emph{DARA\/} project number 50--OR--9501B.
\newline
   We are indebted to G. Gavazzi for providing us with optical data
   prior their publication.

\end{acknowledgements}

\vskip 0.5truecm
\section*{Figure Captions}

{\bf Fig. 1.} Comparison of the observed central CO(1-0) line intensities
(in $\rm K~km~s^{-1}$ in the main beam temperature scale), for 28 VCC galaxies
observed by Stark et al. (1986) (100\arcsec~HPBW) and by Kenney \& Young
(1988) (45\arcsec~HPBW), $I_{\rm CO}(100)$ and $I_{\rm CO}(45)$, respectively.
Both $3 \sigma$ detections and upper limits are shown.
Here, the long-- and short--dashed lines reproduce the relationships
between the couples of measurements in case of point-like and extended
CO distributions with respect to the previous two beam--sizes, respectively.
The filled circles represent the 11 galaxies in common with the present study.
\newline
{\bf Fig. 2.} The observed central [CII] line intensity,
$I_{\rm C II}$, vs. the coverage factor, CF, defined as the ratio
between the area of the observing beam and the projected optical galaxy area
as defined in the RC3.
Hereafter, asterisks and filled triangles denote starburst galaxies
and gas--rich galaxies of the KAO sample, respectively, while open circles
identify spiral galaxies of the ISO sample (see text).
Arrows define upper limits.
\newline
{\bf Fig. 3.} The average column density of molecular hydrogen
in the beam, $2 N$($\rm H_2$), (in terms of HI nuclei)
({\bf a}) and the average column density of neutral atomic hydrogen
in the beam, $N$(HI), ({\bf b}) vs. the coverage factor CF.
See text for the determination of $2 N$($\rm H_2$).
$2 N$($\rm H_2$) and $N$(HI) are not deprojected
from the inclination of the galaxy.
\newline
{\bf Fig. 4.} The relationship between the observed central
[CII] line intensity, $I_{\rm C II}$, and the central
CO line intensity, $I_{\rm CO}$, as derived in Sect. 2.2,
not deprojected from the inclination of the galaxy.
Here, the long-- and short--dashed lines show the average ratios
of the two observables obtained by Stacey et al. (1991) for
the starburst galaxies and the gas--rich galaxies, respectively.
\newline
{\bf Fig. 5.} The rate of [CII] line emission per interstellar hydrogen
atom, $\Lambda \rm = 4 \pi$/$N$(HI), vs. the CF.
Here, the long-- and short--dashed lines show the rates of [CII]
line emission per H atom found by Madden et al. (1993) for the diffuse ISM
of the actively star--forming galaxy NGC\,6946 and by Bennet et al. (1994)
for the high latitude regions of our Galaxy, respectively (see text).
Instead, the two dotted lines delimit the range of values of the rate
of [CII] line emission per H atom in the cold neutral medium
(CNM), as calculated by Wolfire et al. (1995).
(N.B.: the lower dotted line is almost indistinguishable from
the short--dashed line.)
\newline
{\bf Fig. 6.} The rate of [CII] line emission per interstellar hydrogen
nucleus, $\Lambda^{\prime} \rm = 4 \pi$/($N$(HI)$+2 N$($\rm H_2$)),
vs. the CF.

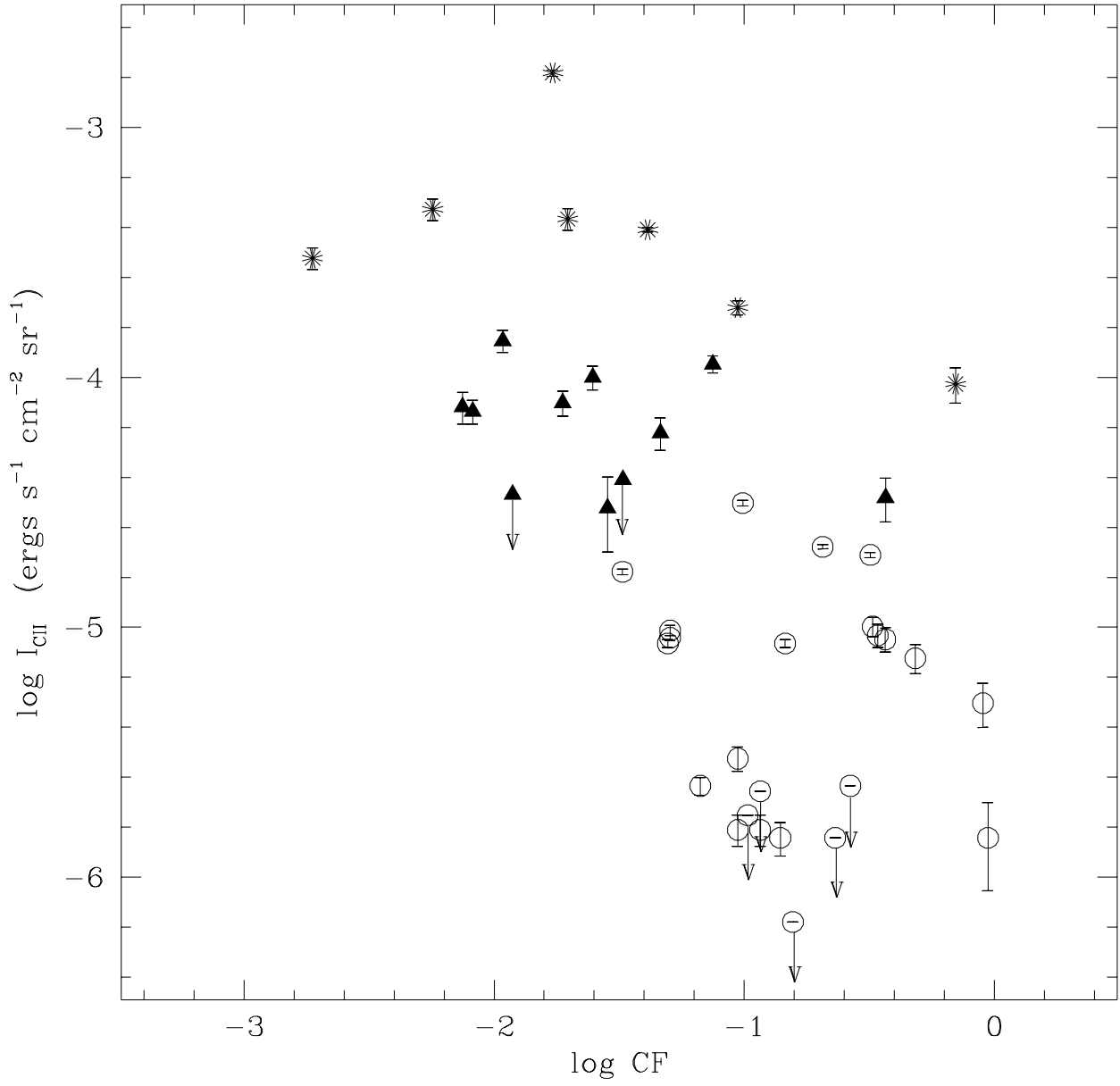
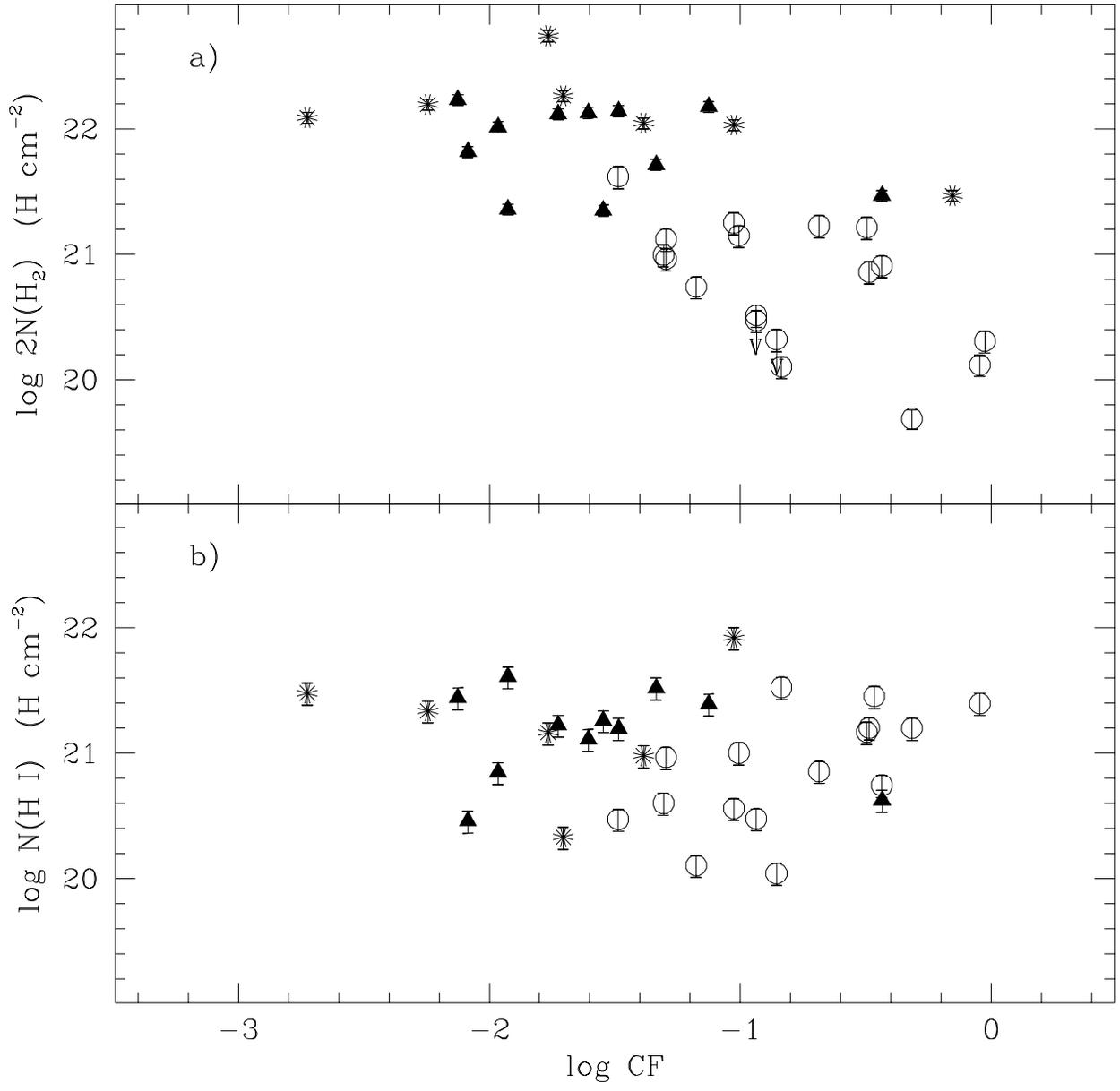
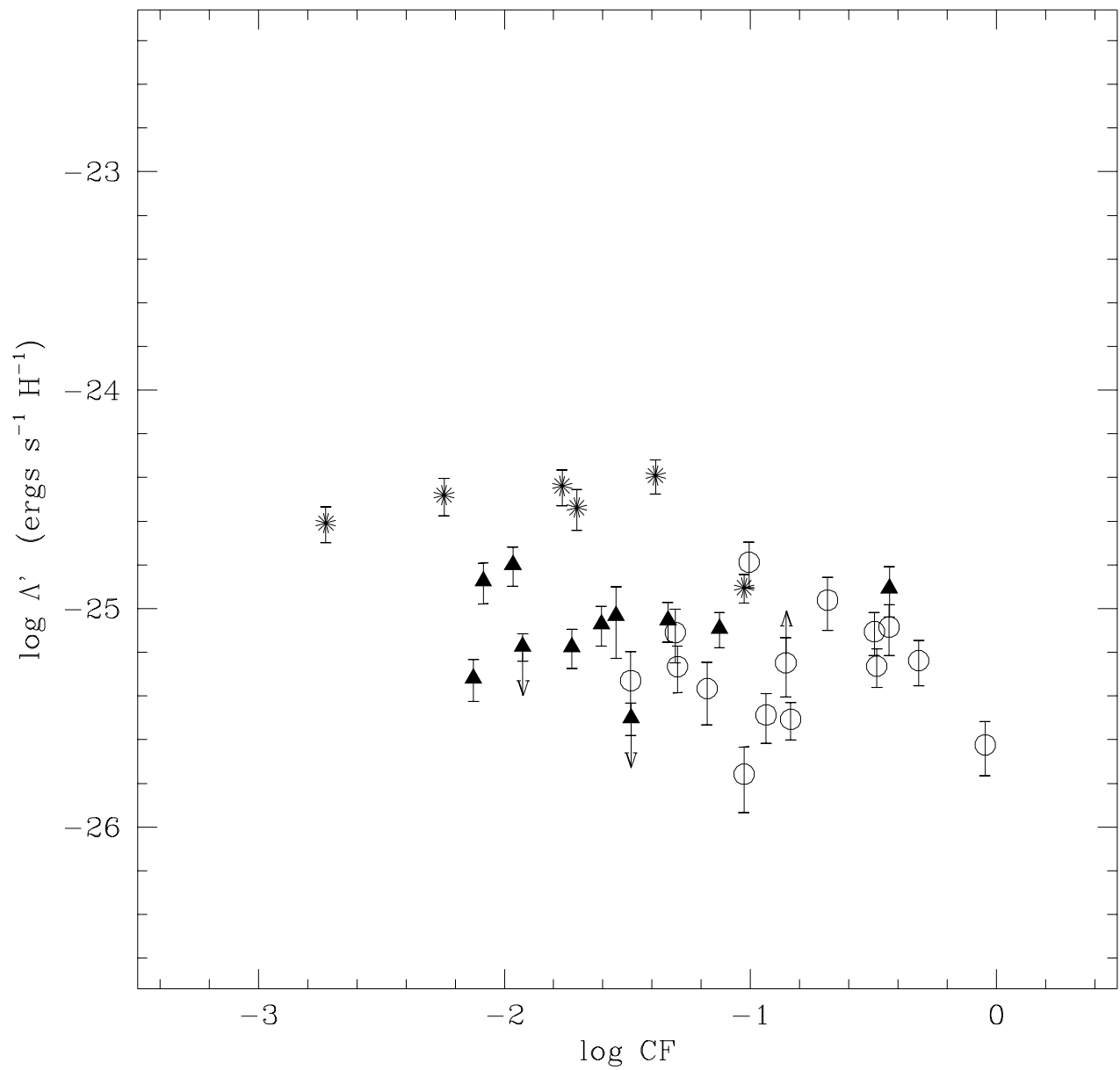
\end{document}